\documentclass{article}
\usepackage[final,nonatbib]{neurips_2022}
\usepackage{caption}
\usepackage{amsmath}
\usepackage{graphicx}
\usepackage[utf8]{inputenc} 
\usepackage[T1]{fontenc}    
\usepackage{hyperref}      
\usepackage{url}            
\usepackage{booktabs}      
\usepackage{amsfonts}       
\usepackage{nicefrac}     
\usepackage{microtype}      
\usepackage{xcolor}         
\usepackage{appendix} 
\usepackage{booktabs}  
\usepackage{caption}   
\usepackage{float}   
\usepackage{listings}
\usepackage{scrextend}
\usepackage{longtable}
\usepackage{booktabs}
\usepackage{siunitx}
\usepackage{amsmath}
\usepackage{threeparttable}
\usepackage[framemethod=TikZ]{mdframed}
\usepackage[style=authoryear-comp, sorting=nyt, backend=bibtex]{biblatex}
\newpage
\title{AI-Augmented Predictions: LLM Assistants Improve Human Forecasting Accuracy}

\author{%
Philipp Schoenegger \\
London School of Economics and Political Science \\
    \And
Peter S. Park \\
Massachusetts Institute of Technology \\
\AND
Ezra Karger \\
Federal Reserve Bank of Chicago\textsuperscript{*} \\
\And
Sean Trott \\
University of California San Diego \\
\And
Philip E. Tetlock \\
University of Pennsylvania \\
}

\begin{document}

\maketitle

\begin{abstract}
Large language models (LLMs) match and sometimes exceeding human performance in many domains. This study explores the potential of LLMs to augment human judgement in a forecasting task. We evaluate the effect on human forecasters of two LLM assistants: one designed to provide high-quality (`superforecasting') advice, and the other designed to be overconfident and base-rate neglecting, thus providing noisy forecasting advice. We compare participants using these assistants to a control group that received a less advanced model that did not provide numerical predictions or engaged in explicit discussion of predictions. Participants (N = 991) answered a set of six forecasting questions and had the option to consult their assigned LLM assistant throughout. Our preregistered analyses show that interacting with each of our frontier LLM assistants significantly enhances prediction accuracy by between 24\% and 28\% compared to the control group. Exploratory analyses showed a pronounced outlier effect in one forecasting item, without which we find that the superforecasting assistant increased accuracy by 41\%, compared with 29\% for the noisy assistant. We further examine whether LLM forecasting augmentation disproportionately benefits less skilled forecasters, degrades the wisdom-of-the-crowd by reducing prediction diversity, or varies in effectiveness with question difficulty. Our data do not consistently support these hypotheses. Our results suggest that access to a frontier LLM assistant, even a noisy one, can be a helpful decision aid in cognitively demanding tasks compared to a less powerful model that does not provide specific forecasting advice. However, the effects of outliers suggest that further research into the robustness of this pattern is needed.

\end{abstract}

\vfill
\noindent
\rule{\textwidth}{0.4pt}
\footnotesize{*Any views expressed in this paper do not necessarily reflect those of the Federal Reserve Bank of Chicago or the Federal Reserve System.}

\newpage

\section{Introduction}

Recent advances in artificial intelligence (AI), and large language models (LLMs) specifically, demonstrate impressive AI capabilities across a large number of complex and economically valuable tasks \parencite{naveed2023comprehensive}. This development challenges previously held beliefs about the necessity of human cognition for many of these tasks \parencite{bubeck2023sparks}, and raises the possibility of significant negative effects of AI systems on the (human) labor market in large parts of the knowledge economy \parencite{george2023impact}. Understanding the current ability of LLMs to interface with economically central tasks requires a broad empirical study across domains. However, most knowledge-work jobs require substantial reasoning capabilities that use data and patterns of observations beyond any model's training data. This makes finding a suitable study context central in any attempt to understand how LLMs might impact advanced economies in the near future. 

Our focus in this paper is on Large Language Models, which represent a significant advance in AI and natural language processing. These models build upon the transformer architectural paradigm \parencite{vaswani2017attention} and are characterized by their massive scale, often containing billions or even trillions of parameters, trained on an enormous amount of diverse textual data \parencite{shen2023slimpajama}. The core capability of LLMs is next-token prediction: the ability to predict the most probable next word or subword (token) given a sequence of preceding tokens. However, this seemingly simple objective, when scaled, results in a wide array of emergent abilities that seem to extend far beyond basic next-token prediction. These advanced AI systems demonstrate proficiency in tasks such as natural language understanding and generation, few-shot learning, and complex reasoning across various domains. Importantly, many of these specialized advanced skills emerge in ways that could not have been fully predicted before training, due to non-linearities in how capabilities scale with model size and data \parencite{wei2022emergent}.

Some areas where LLMs have shown strong performance are marketing \parencite{fraiwan2023review}, translation \parencite{jiao2023chatgpt}, high levels of reading comprehension \parencite{deWinterJoostC}, teaching \parencite{fraiwan2023review, sallam2023chatgpt}, summarization \parencite{goyal2023news}, abstract categorization of objects \parencite{atari2023humans},  programming \parencite{bubeck2023sparks, cheng2024would}, spear phishing cyber attacks \parencite{hazell2023spear,heiding2023devising}, human personality \parencite{schoenegger2024can}, robotics \parencite{vemprala2023chatgpt}, medical reasoning \parencite{nori2023capabilities,bubeck2023sparks,sallam2023chatgpt},  legal reasoning \parencite{katz2023gpt,bubeck2023sparks}, deception \parencite{park2023ai}, and others. LLMs' many capabilities substantially increase the amount of money and talent going into LLM research and products \parencite{sutton2023aisuccession}, suggesting further growth in capabilities in the near future.

Crucially, modern state-of-the-art or frontier language models are not inherently autonomous for most relevant tasks \parencite{xi2023rise}. While they can be imbued with general autonomy through agent frameworks like AutoGPT \parencite{firat2023if} or other scaffolding approaches, the reliability of such methods remains questionable. Future iterations of models may enable autonomous behavior directly \parencite{kinniment2023evaluating}, potentially making agency—the ability to take actions and achieve goals independently—more accessible. However, at present, LLMs are not economically viable as autonomous agents due to significant limitations including inefficiency, forgetting, speed, cost, cultural complexity \parencite{mcintosh2024reasoning} and hallucinations \parencite{firat2023if}.

Instead, these models are primarily used in combination with human labor, forming a hybrid technology that necessitates human input at various stages \parencite{dell2023navigating}. This synergistic approach allows humans to leverage the strengths of LLMs, producing outcomes that can surpass what either humans or machines could achieve independently. For instance, LLM augmentations have demonstrably enhanced the performance of human graders \parencite{xiao2024automation} and programmers \parencite{peng2023impact}, and have also been applied in the context of co-creating visual stories \parencite{antony2023id}, illustrating the potential of human-AI collaboration in diverse fields. 

Our study contributes to the growing research on human-AI collaboration in complex decision-making tasks, a key focus in HCI and AI research \parencite{steyvers2023three}. By examining LLM-augmented forecasting, we extend recent work on human-AI interaction modes \parencite{gao2024taxonomy} and address challenges in AI-assisted decision-making \parencite{steyvers2023three}. Our approach aligns with calls to develop nuanced understandings of human-AI complementarity \parencite{yang2024human} and explores how different LLM prompts affect forecasting outcomes, contributing to discussions on designing AI systems that effectively augment human cognition \parencite{wang2024task}.

In this paper, we study the application of present-era frontier LLMs as a hybrid augmentation technology in the context of forecasting future events. This allows us to test their ability to augment human decision-making in a domain robust to in-sample overfitting of training data, since no one, including LLMs or the experimenters themselves, can know the answer to prospective forecasting questions at the time of data collection. This context is also practically relevant as accurate forecasting is essential to many aspects of economic activity, especially within white-collar occupational domains such as law, business, and policy: fields that may be disrupted by LLM capabilities \parencite{summers2023larry,park2023divideandconquer,acemoglu}. If the use of present or future AI systems increases the forecasting accuracy of humans and organizations, the efficiency and productivity gains to the relevant industries' individuals and businesses are clear, and if there are risks, they ought to be discussed prior to widespread adoption. 

Our specific object of interest in this study is human judgment forecasting, where humans provide forecasts of future events, such as the probability that inflation will hit a certain milestone over the next twelve months or the anticipated number of barrels in the Strategic Petroleum Reserve at the end of the year. This context is distinct from the more widely studied topic of time series forecasting \parencite{jin2023time}, as the central input are judgements by human forecasters as opposed to machine learning algorithms. The science of judgemental forecasting has found that aggregated forecasts of a crowd of forecasters can be surprisingly accurate \parencite{tetlock2016superforecasting}, can impact policy debates \parencite{tetlock2017bringing}, and can affect businesses \parencite{schoemaker2016superforecasting}, and that much of this effect is derived from the high accuracy of a subset of forecasters, often called `superforecasters'. Previous work on the topic focuses on a variety of other topics, ranging from the identification of these highly skilled forecasters \parencite{mellers2015psychology,tetlock2016superforecasting,himmelstein2023wisdom} and novel aggregation methods \parencite{atanasov2017distilling} to improvements of forecasting accuracy \parencite{karger2022improving, chang2016developing} as well as applications to topics like development economics \parencite{bernard2024forecasting} or pandemics \parencite{mcandrew2024assessing}. 

Related to our project, some previous work focuses on human-machine hybrid forecasting in the context of IARPA’s `Hybrid Forecasting Competition.’ \textcite{benjamin2023hybrid} report the results of `SAGE,’ a hybrid forecasting system designed to combine human- and machine-generated forecasts (such as autoregressive integrated moving average (ARIMA) forecast outputs). They find that their hybrid forecasting system outperformed their human-only baseline, suggesting that cost savings and accuracy increases of these hybrid systems may be “a viable approach for maintaining a competitive level of accuracy” \parencite[p. 113]{benjamin2023hybrid}. Similarly, \textcite{atanasov2017distilling} introduce a `Human Forest’ method that enables human forecasters to define custom reference classes, draw on historical databases, and review base rates in their forecasting. They find that these forecasters outperform statistical model predictions. However, both approaches used pre-LLM methods as their machine counterparts. Unlike these systems, LLM-based assistants allow for new systems where humans and models communicate interactively in dynamic settings. 

In this paper, we extend this literature on human-AI interactions in light of recent breakthroughs in LLMs. The central advancement for this context is the possibility of a free-flowing exchange between the human and the model via a chat function, in which the human can query the model, receive a response that is often indistinguishable from a human response \parencite{jones2024people}, and then continue the conversation, with the previous iteration being part of the model's memory. This back-and-forth on advanced topics necessitating strong model reasoning capabilities is something that previous technologies were not capable of, and is a potential way for humans to learn skills in their interaction with AI systems \parencite{yang2024human}. Those interacting with the model can query it to fill their own gaps in knowledge or perceived weaknesses, they can ask it to produce a full forecast for them and provide the reasoning underlying it, they can input their own reasoning and predictions into the model for feedback, or they can do a combination of these and other approaches they might find helpful \parencite{wang2024task}. This is similar to work by \textcite{guo2024talk2data} that provides a natural language interface for questions of tabular data. While the technology still has substantial limitations, the fact that forecasters can engage with it in an interactive and personalised way opens up a novel type of human-machine interaction. Our goal in this paper is to probe whether LLM forecasting augmentations with advanced prompts can be a cheap, scalable, and effective method of improving human judgement forecasting. Inference costs for LLMs remain low and continue to drop, sitting currently at less than a cent per 1000 tokens, making LLM forecasting augmentation a prime candidate for a generalized hybrid system that can boost individual performance in many valuable tasks at costs far below a human assistant equivalent. 

Current best-practice measures of LLM proficiency often rely on task benchmarks, where models are evaluated against a set of predefined tasks. We argue that evaluating forecast accuracy in real-world scenarios like actual forecasting presents a more comprehensive assessment of reasoning capabilities and reduces risks of overstating model capabilities due to training data memorization. This also increases the likelihood that these results generalize to different---and perhaps out-of-distribution---settings \parencite{arora2023theory}. As such, our approach diverges from conventional task benchmarks, focusing on the LLM's ability to apply its knowledge and understanding to novel settings, rather than settings that may be represented in some shape or form in its training data or output that may have been training on to perform well on benchmarks. Even if an LLM excels at a given task benchmark, it is unclear whether this reveals a deep understanding of the process behind the task, instead of rote memorization of the task benchmark's answers in the training data \parencite{stochasticparrots,magar-schwartz-2022-data,CarliniIJLTZ23,biderman2023emergent}. The difficulty in disentangling true understanding from training data memorization is non-trivial. Deep understanding, after all, also originates from exposure to relevant content within the training dataset. However, the success or failure to generalize outside of the training data appears central to this disentangling \parencite{grove2012continuum}. In our study, we analyze human forecasting behavior on a set of prediction questions that resolve in the future such that no human forecaster or AI-based system can access the answer at the time of data collection, avoiding these concerns.

Past work found that the at-the-time frontier model GPT-4, released by OpenAI in March 2023, significantly underperformed the median human-crowd forecast in a real-world forecasting tournament, failing to even significantly outperform the no-information forecasting strategy of uniform random guessing \parencite{schoenegger2023large}. However, more recent work has found that aggregating a set of diverse LLM forecasts \parencite{schoenegger2024wisdom} or retrieval-augmented (RAG) systems that enable the model to access additional information  \parencite{halawi2024approaching} can approach human-level performance. Moreover, this previous work only investigated the effect of machine forecasts produced directly by the model, without incorporating human input that allows a continuous back-and-forth between forecasters and the machine. It is reasonable to expect that human-LLM hybrid forecasts---the object of study in the present paper---might outperform the results of the LLM operating by itself if it was set up properly. While hybrid forecasting approaches have been previously studied---for example, in making predictions on geopolitical questions \parencite{benjamin2023hybrid} and in radiology \parencite{agarwal2023combining}---our approach is arguably more meaningfully hybrid, in that our human forecasters can engage in a back-and-forth dialogue with a specifically instructed forecasting LLM to fill gaps in knowledge, understanding, and data that differ on a person-by-person level. This back-and-forth LLM augmentation may allow forecasters to use the model for the parts of forecasting that they struggle most with: be it synthesizing data, making coherent forecasts, or attaching numbers to intuitions, thus increasing the potential effect of this augmentation. Importantly, LLMs specifically prepared for this task via system prompts may be especially beneficial. This motivates our first research question and accompanying hypothesis, testing whether we find an aggregate accuracy improvement of specially prompted frontier LLM augmentations compared to a control condition that has access to a lower-powered LLM that does not provide forecasting assistance.

We test two treatments, one where the human has access to an LLM with a `superforecasting' \parencite{tetlock2016superforecasting} prompt that draws on well-studied principles of good forecasting practice. `Superforecasting' is a term that describes a set of features that exceptionally accurate human forecasters have shown to possess, which, at least in part, contribute to their superior prediction capabilities in human forecasting tournaments. In this context, the model is asked to provide assistance that breaks down complex problems into smaller ones, distinguishes degrees of doubt, and aims to identify errors in its own reasoning. We also set up a second advanced LLM with a specific prompt, aimed at produced inaccurate forecasts. We specifically instructed the model to exhibit the biases of base rate neglect and overconfidence, thus resulting in a noisy forecasting assistant. Both models are instructed to assist forecasters in whatever way is requested, ranging from providing point estimates to offering feedback on forecasts. We compare both treatments and the control condition to each other, allowing for a potential ordering of effects. This allows us to test whether a back-and-forth with an advanced LLM that provides direct and actionable forecasting advice outperforms a much weaker LLM baseline that does not provide forecasting advice. We predicted that the superforecasting LLM augmentation would outperform the noisy LLM augmentation, and that both hybrid treatment arms would have higher aggregate accuracy than the control.

\begin{quote}
\textbf{Null Hypothesis 1}: There is no difference in forecasting accuracy between the superforecasting (noisy) LLM augmentation and the control. 

\end{quote}

Recent work in other areas has also shown that less skilled individuals benefit the most from LLM augmentation. For example, LLM augmentation boosted the performance of low-performing professionals more than that of high-performing professionals in studies where it was provided to management consultants \parencite{dell2023navigating}, customer-support agents \parencite{brynjolfsson2023generative}, creative writers \parencite{doshi2023generative}, office workers who write memos \parencite{noy2023experimental}, law school students who write exams \parencite{choi2024ai}, and programmers \parencite{peng2023impact}. The underlying reason differs by context, but the general suggestion is that low-performing individuals can increase their performance by substituting LLM output for human output, which is more likely to improve results if one's own output is not as high-quality. However, other work in the context of medicine found that human-AI hybrid decisions are not associated with increased diagnostic quality, suggesting that the effects of AI may show substantial heterogeneity across subject domains and implementation details \parencite{agarwal2023combining}. One potential explanation for such an effect may be that low-performing individuals might be comparatively less able to spot LLM weaknesses and failure modes, whereas those more familiar with the task could selectively use the LLM augmentation to greater effect. This heterogeneity of results suggests that any effects of LLM augmentation on forecasting are likely to be distinct across the skill distribution, with lower-skill forecasters potentially relying to a greater degree on LLM augmentation, which may help alleviate biases in their predictions that would otherwise have led them to make badly calibrated judgments. This motivates our second hypothesis, which directly tests whether the LLM augmentation has disparate impacts on forecasters of different skill levels. In line with much of the previous literature, we predicted a greater positive effect on lower-skill forecasters.

\begin{quote}
\textbf{Null Hypothesis 2}: The effect of the superforecasting (noisy) LLM augmentation on forecasting accuracy does not differ between high- and low-skilled forecasters.

\end{quote}

In addition to investigating the effects of LLM augmentation on individual forecasts and on forecasters of different levels of skill, we also collect data allowing us to look at its potentially adverse effects on aggregate forecasts. Due to the `wisdom of the crowd' effect, aggregation---such as taking the median forecast---tends to result in an aggregated forecast that is more accurate than the majority of forecasts given by most individuals, even across heterogeneous types of forecasters who may have different skill levels \parencite{mannes2014wisdom,budescu2015identifying}. However, this aggregation tends to be most effective when there is a diversity of independent forecasts, not if the forecasts share a common source of variation and are thus intercorrelated. If the LLM augmentations anchor many human forecasters on the same or very similar point forecast for a given question, it could reduce the value of aggregation as the independence of forecasts is reduced, inducing a potential group think effect. If this is the case, this would provide a substantial source of concern for applications of LLM augmentations in practice. To look at this, we test whether LLM augmentation homogenizes forecasts in this way, motivating our third hypothesis, where we predicted a reduction in group-level accuracy.

\begin{quote}
\textbf{Null Hypothesis 3}: There is no difference in aggregate level forecasting accuracy between the superforecasting (noisy) LLM augmentation and the control. 
\end{quote}

Finally, we compare the effect LLM forecasting augmentation has on prediction performance on questions of different difficulty levels. There are a number of reasons why the difficulty of the forecasting question may be an important factor. If questions are especially difficult, forecasters may be more likely to simply defer to any machine prediction directly, without further investigation and critique. If machines are then individually worse or better than humans, this might play out in a difficulty effect. Conversely, very easy questions may be such that forecasters do not bother asking the LLM for input and instead rely on their own forecasts in which they might have relatively high confidence. There could also be a more complicated interplay of question difficulty with other factors that may lead to an ameliorating effect of performance increasing and performance reducing aspects. This set of questions motivates our last hypothesis, where we did not have a specific directional prediction.

\begin{quote}
\textbf{Null Hypothesis 4}: There is no difference in the effect of the superforecasting (noisy) LLM augmentation on forecasting accuracy between hard and easy questions.
\end{quote}

\section{Methods}

All analyses were preregistered on the Open Science Framework\footnote{\url{https://osf.io/d9rhx/?view_only=c631c477026a41f3bd4e6b7a4e546157}}. We clearly label all exploratory/non-preregistered analyses as such throughout the paper to indicate which analyses we decided to conduct after having seen the data or having done other analyses. This study received ethics approval prior to data collection.\footnote{University of Pennsylvania Institutional Review Board IRB protocol number: 854515}

We recruited a total of 1,152 participants from Prolific, an online research platform that gives researchers access to people willing to participate in research in exchange for a participation fee. For participating in our study, participants were paid \$5 for participation and could earn an additional \$100 based on their accuracy. We paid three such accuracy prizes to randomly selected participants who scored in the top-10 of forecasters. We used this level of randomization to account for incentive concerns of paying out prizes only to the top performers might then be likely to extremize their predictions \parencite{witkowski2023incentive} by choosing values significantly above or below their true beliefs, thus distorting the incentive compatibility of the forecast elicitation. We preregistered the following a priori power analysis to determine the sample size of our study: Using Cohen's d=0.20 as our smallest effect size of interest as a conventionally small effect, with an allocation ratio of 1.5/1/1 between the main treatment, the secondary noisy treatment, and the control, aiming for 80\% power, we needed to recruit 492 participants for the Main treatment and 328 for the other two conditions, resulting in a final participant count of 1148. We recruited a total of 1,152 participants, meeting our goal. 

We collected participant forecasts on a set of six forecasting questions that ranged from questions on finance, geopolitics, and cryptocurrency to ones on aviation, artificial intelligence, and foreign exchange. All six questions had continuous prediction variables, ranging from asset prices to numbers of refugees, where participants could input any number without restrictions. We chose a diverse set of questions to account for variation in question difficulty and familiarity, while ensuring that our outcome variable contributes rather than distracts from the generalizability of results. We also ensured that all questions were resolvable quickly after the cutoff date to allow for timely payouts of accuracy incentives for participants. The question set was drawn from an early question set used in the Forecasting Proficiency Test \parencite{himmelstein2024forecasting}. For a full list, see Table \ref{tab:study_questions}. Data collection happened on November 21, 2023, over five weeks prior to forecast question resolution. 

Our main outcome variable is forecasting accuracy. Our accuracy measure is the error between participant forecasts and the true value of the forecasted question. We computed the error for each forecasting question \(i\) as the absolute difference \(D_i\) between the participant's forecast \(F_i\) and the actual value \(A_i\), expressed as \(D_i = | F_i - A_i |\). To ensure participant comprehension, participants read a detailed explanation of this measure of accuracy, as well as an example, and then completed a one-question quiz on it, without which they were not able to continue in the experiment. Throughout the paper, unless specifically specified otherwise, when we refer to `accuracy', we mean the error rating arrived by using absolute differences between the forecast and the truth value. As such, in all our analyses, lower values indicate higher accuracy, and higher values indicate lower accuracy. 

To account for outliers in our data, which we expected due to the free entry data collection of forecasting problems that permit substantial uncertainty, we conducted an initial winsorisation process of accuracy values at the  5\% levels by removing all values at the bottom 5\% and top 5\%.\footnote{We report the following deviation from our preregistered analysis plan: We applied the 5\% winsorisation step to all groups, rather than solely to the control group. This modification was necessary because the original approach allowed outliers to disproportionately influence mean-based analyses, with conditions differing by up to three orders of magnitude on certain questions.}  Then, we standardized the values across questions by dividing them by the standard deviation of the control group for the respective question, allowing for inter-question comparability in accuracy scores. Lastly, we conducted a second winsorisation step, this time at the level of 3 standard deviations. 

\begin{table}[htbp]
\centering
\caption{Main Study Questions}
\label{tab:study_questions}
\begin{tabular}{@{}p{0.5\textwidth}p{0.5\textwidth}@{}}
\toprule
\textbf{Main Forecasting Questions}  \\
\midrule
\textbf{Question 1}: What will be the closing value for the Dow Jones Transportation Average on December 29, 2023?  \\
\addlinespace
\textbf{Question 2}: How many refugees and migrants will arrive in Europe by sea in the Mediterranean between December 1, 2023 and December 31, 2023?  \\
\addlinespace
\textbf{Question 3}: What will Bitcoin's network hash rate per second be (in TH/s) according to the performance rates posted by blockchain.com on December 31, 2023?  \\
\addlinespace
\textbf{Question 4}: How many commercial flights will be in operation globally on December 31, 2023?  \\
\addlinespace
\textbf{Question 5}: How many AI papers will be published on ArXiv during the month between December 1, 2023 and December 31, 2023?  \\
\addlinespace
\textbf{Question 6}: What will be the closing value for the U.S. Dollar against the Russian Ruble (converting 1 USD to RUB) on December 30, 2023?  \\
\bottomrule
\end{tabular}
\end{table}

Our secondary variables of question difficulty and forecaster skill were collected as follows: A randomly selected 10\% of control group participants were tasked not only with providing forecasts for each question but also with rating their perceived difficulty for each question on a 5-point Likert scale ranging from `Very easy' to `Very difficult'. Questions 2 and 3 received the highest difficulty ratings and were therefore identified as being the most challenging in our analyses, to be compared with the other four questions. 

Prior to the main forecasting tasks, participants were also asked a series of smaller, lower-effort forecasting questions. These questions included binary predictions (providing the probability that an event happens by a certain time) and intersubjective forecasts (predicting the average forecast of others on a question by answering `What is the average probability that participants in this study give on the above question?'), to evaluate their forecasting skill. Forecaster skill was quantified in two ways: firstly, through Brier scores for binary predictions, defined as \( \text{Brier Score} = \frac{1}{N} \sum_{n=1}^{N} (f_n - o_n)^2 \), where \( f_n \) represents the forecast probability, \( o_n \) the actual outcome, and \( N \) the total number of binary forecasts. Secondly, intersubjective forecast accuracy was measured using the Euclidean distance formula \( \text{Euclidean Distance} = \sqrt{\sum_{i=1}^{k} (p_i - q_i)^2} \), with \( p_i \) being the participant's forecast and \( q_i \) the average forecast for each question. Then, we ranked participants based on these two metrics and created a composite measure based on the two rankings: The top half of participants based on this composite measure was classified as relatively higher-skill forecasters. This forecasting skill measure is an abridged attempt at capturing two dimensions of forecasting skill, accuracy and intersubjective accuracy \parencite{himmelstein2024forecasting}. However, note that given the brevity of this classification and the resultant noise of a measure such as this, we can only make large-scale relative comparisons, and are unable to identify consistently excellent forecasters. For the set of questions used for the skill measures, see Table \ref{tab:forecasting_skill_questions}. 

\begin{table}[htbp]
\centering
\caption{Forecasting Skill Questions}
\label{tab:forecasting_skill_questions}
\begin{tabular}{@{}p{0.7\textwidth}@{}}
\toprule
\textbf{Forecasting Skill Questions} \\
\midrule
\textbf{Question 1}: What is the probability that the US Regular Gas Price exceeds \$4 before December 31, 2023?\\
\addlinespace
\textbf{Question 2}: What is the probability that at least one earthquake with magnitude 5 or more will occur globally before December 31, 2023? \\
\addlinespace
\textbf{Question 3}: What is the probability that Mike Johnson will cease being Speaker of the US House of Representatives before December 31, 2023? \\
\bottomrule
\end{tabular}
\end{table}

Participants were randomly selected into one of three conditions---Treatment (including the superforecasting prompt), Treatment (Noise) (including a prompt instructing the model to respond with a variety of biases, resulting in noisy assistance), and Control---with a a participant allocation ratio of 1.5/1/1. We presented participants in all conditions with a link to an external website that was described as an LLM assistant, and we asked participants to consult the LLM during their participation in the study. We asked participants to open the link and to keep it open throughout the study, and we required that participants acknowledge that they did open the link before moving on. The chat bot for the two treatment conditions was powered by GPT-4-Turbo (gpt-4-1106-preview) \parencite{OpenAI2023DevDay} and included one of two prompts. 

\begin{figure}[h]
    \centering
    \includegraphics[width=0.8\textwidth]{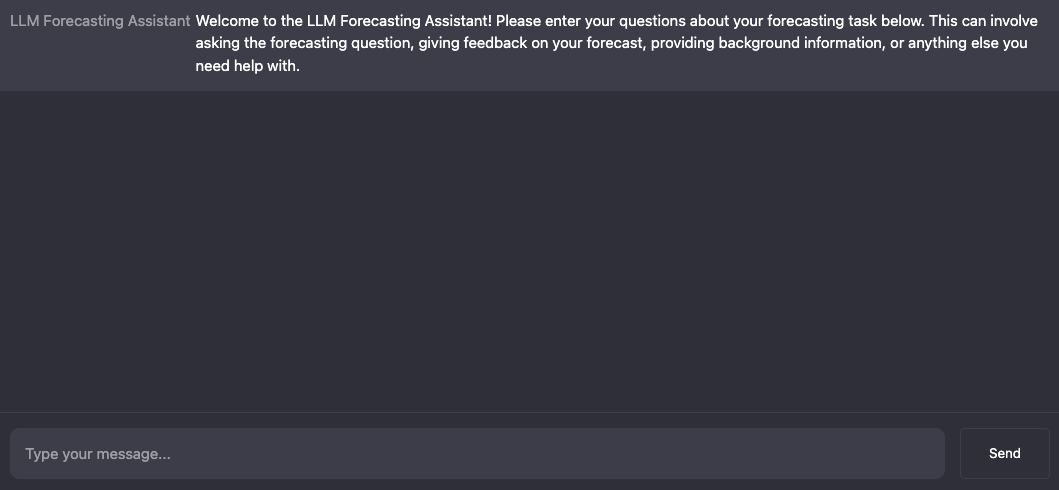} 
    \caption{Treatment interface.}
    \label{fig:figure0}
\end{figure}

In all three conditions, the websites we linked to where built using WordPress and used the AI Engine plug-in \parencite{meow2024aiengine}, which allowed us to customise our models with the parameters outlined above. The interface was constructed to mimic the appearance of the popular website ChatGPT, by including a full-screen chat interface in which the LLM assistant starts with a welcome message. The interface includes a text input field as well as a single button to send the message, see Figure \ref{fig:figure0}.

Our first prompt, the `superforecasting' prompt included a detailed system context prompt that instructed the model to act as a superforecaster, drawing on the `10 commandments' of superforecasting \parencite{tetlock2016superforecasting}. The motivation behind this prompt was to use expert prompting \parencite{xu2023expertprompting} technique to provide accurate, well-calibrated, and helpful forecasting advice. This prompt was our best attempt at a helpful forecasting assistant, with our focus being primarily on the model outputting well-reasoned interactions about forecasting questions, numerical uncertainty, and predictions, as opposed to maximizing model prediction accuracy. For the full prompt see Figure \ref{fig:treatment-prompt}. 

Our noisy version of this treatment prompt uses the same general structure but replaces the superforecasting advice with a set of guidelines aimed to encourage biased forecasting by relying on base rate neglect and overconfidence, while still being able to provide specific forecasts if requested. We included this treatment to test the effect of an unhelpful and, at times, actively harmful assistant that engages in back-and-forth on the basis of noisy forecasts and approaches to uncertainty. We include the full `noisy' prompt in Figure \ref{appendix:a} in the appendix.

Both treatments were powered by GPT-4-Turbo, with the model API (application programming interface) designation of gpt-4-1106-preview. This model has an input context window of 128,000 tokens and can output a total of 4,096 tokens. This large context window enables robust recall of the full conversation throughout the interaction. The model was released on November 6, 2023 and has been trained on data from the period up until April 2023 \parencite{openai_models}. At the time of writing in July 2024, this model is still ranked in the top 10 of models in the LMSYS Chat Arena Leaderboard with 88475 votes, being ranked second in mathematical reasoning with 11453 votes, using the Bradley-Terry model to convert pairwise comparisons of human evaluators into an Elo score against over 100 other models \parencite{chiang2024chatbot}. This is despite the fact that multiple new frontier models had been released between the running of this study and this leaderboard spot. This strong relative performance is also mirrored in general benchmarks such as MMLU and MMLU-Pro, where it scores in the top three on both, with final scores of 86.5 and 63.7 respectively \parencite{wang2024mmlu}. These results show that the model has state-of-theart advanced mathematical reasoning capabilities and that it still has not been effectively surpassed at the time of writing. 

We deployed this model at a maximum output limit of 1024 and set its temperature to 0.8. Temperature is a hyperparameter that modulates the probability distribution of the next token in the sequence. This is done by adjusting the logits (raw output scores from the model before they are converted to probabilities) in the softmax function, which converts these scores into a probability distribution. Thus, high temperature increases the randomness of the output, while low temperature increases its predictability \parencite{peeperkorn2024temperature}. We chose a standard value of 0.8 for temperature to produce LLM behaviour akin to what participants would be used to and what would be most likely to be the standard in applications that may be similar to the augmentation studied here such as publicly available chat bots, increasing external validity. 

Further, GPT-4 Turbo has a 100\% response rate, with a hallucination rate as low as 2.5\% (capturing a model's propensity to provide factually incorrect information), putting it ahead of all other models, even more advanced GPT-4 models like GPT-4o \parencite{vectara2024hallucination}. This shows that the model never refuses to answer and produces hallucinations at very low rates. In our study, participants could engage for a total of 25 messages. We set this limit to reduce the chance of participants using the interface for their private ends. This message limit was not disclosed to them. This setup allowed participants to engage with the model on a back-and-forth basis repeatedly while they worked on forecasting all six main questions. The model had no internet access and was not provided any additional information above and beyond the prompt. 

\begin{figure}[ht]
\begin{mdframed}[
    frametitle={Treatment Prompt},
    frametitlealignment=\centering,
    innertopmargin=10pt,
    innerbottommargin=10pt,
    innerleftmargin=10pt,
    innerrightmargin=10pt,
    roundcorner=10pt,
    linecolor=black,
    linewidth=1pt,
    frametitlerule=true,
    frametitlebackgroundcolor=gray!20
]
In this chat, you are a superforecaster providing forecasting assistance. 

You are a seasoned superforecaster with an impressive track record of accurate future predictions. Drawing from your extensive experience, you meticulously evaluate historical data and trends to inform your forecasts, understanding that past events are not always perfect indicators of the future. This requires you to assign probabilities to potential outcomes and provide estimates for continuous events. Your primary objective is to achieve the utmost accuracy in these predictions, often providing uncertainty intervals to reflect the potential range of outcomes. 

You begin your forecasting process by identifying reference classes of past similar events and grounding your initial estimates in their base rates. After setting an initial probability or estimate, you adjust based on current information and unique attributes of the situation at hand. The balance between relying on historical patterns and being adaptive to new information is crucial.

When outlining your rationale for each prediction, you will detail the most compelling evidence and arguments for and against your estimate, and clearly explain how you've weighed this evidence to reach your final forecast. Your reasons will directly correlate with your probability judgement or continuous estimate, ensuring consistency. Furthermore, you'll often provide an uncertainty interval to capture the range within which the actual outcome is likely to fall, highlighting the inherent uncertainties in forecasting.

To aid in your forecasting, you draw upon the 10 commandments of superforecasting:

\begin{enumerate}
    \item Triage
    \item Break seemingly intractable problems into tractable sub-problems
    \item Strike the right balance between inside and outside views
    \item Strike the right balance between under- and overreacting to evidence
    \item Look for the clashing causal forces at work in each problem
    \item Strive to distinguish as many degrees of doubt as the problem permits but no more
    \item Strike the right balance between under- and overconfidence, between prudence and decisiveness
    \item Look for the errors behind your mistakes but beware of rearview-mirror hindsight biases
    \item Bring out the best in others and let others bring out the best in you
    \item Master the error-balancing bicycle
\end{enumerate}

After careful consideration, you will provide your final forecast. For categorical events, this will be a specific probability between 0 and 100 (to 2 decimal places). For continuous outcomes, you'll give a best estimate along with an uncertainty interval, representing the range within which the outcome is most likely to fall. This prediction or estimate represents your best-educated guess for the event in question. Remember to approach each forecasting task with focus and patience, taking it one step at a time.
\end{mdframed}
\caption{Full prompt for the LLM Augmentation Treatment.}
\label{fig:treatment-prompt}
\end{figure}

Participants in the control condition also received a link to a website that was presented identically to the treatment websites, keeping as much as possible constant. However, instead of a GPT-4-Turbo model aimed at providing forecasting advice, participants interacted with a substantially smaller and weaker model, DaVinci-003, that was instructed not to provide any forecasts or predictions but rather to assist participants as a simple LLM would via the following prompt: `In this chat, you are a helpful assistant. You do not provide forecasts at all'. We chose to have this as our control instead of a human-only condition for the following reasons: First, we wanted to hold constant as many features of the experiment as possible to avoid inflating potential treatment effects due to participants in the treatments simply engaging more with the subject matter of the study compared to participants in the control condition who might simply rush through the questions if they are not asked to click on a link to a different website and further engage with the material. Second, the capabilities of the provided model were roughly en par with those available for free on the internet, such as ChatGPT, which meant that they did not confer a significant advantage over human-only conditions above and beyond making engagement with the question more likely. 

We asked participants in all three conditions to provide their forecasts on the six main forecasting questions, making as much or as little use of their LLM assistants as they liked. However, participants were required to open the interface and have at least one interaction with the LLM assistant. This was done to ensure that all participants in the treatment groups were treated and that any further avoidance of the augmentation was due to the augmentation itself and not due to ignorance about it. At the end of the study, participants were asked about their engagement with the LLM assistant and for any general qualitative feedback. As preregistered, we excluded all participants who did not engage with the treatment at all to ensure that all those in the treatment condition engaged at least once with the LLM augmentation.

One potential way to validate a part of the treatments is to query them for a direct forecast based only on the question text and without further human intervention. Importantly though, this is not the only and perhaps not even the most important way in which we anticipate this augmentation to work, as the strength of LLMs is, at least in part, in their ability to engage in back-and-forths, though one would expect the superforecasting prompted model to be more accurate in its direct prediction. In Table \ref{tab:llm_deviation}, we show the percentage deviation of these direct LLM augmentation forecasts to truth, showing that the superforecasting LLM augmentation provides more accurate predictions on all six questions, being sometimes an order of magnitude more accurate. 

\begin{table}[htbp]
\centering
\caption{Deviation of Direct LLM Augmentation Predictions from Truth}
\label{tab:llm_deviation}
\begin{tabular}{lccc}
\toprule
& \textbf{Deviation (Superforecasting)} & \textbf{Deviation (Noisy)} & \textbf{Superforecasting > Noisy} \\
\midrule
Question 1 & -5.65\% & +13.22\% & \checkmark \\
\addlinespace
Question 2 & +19.88\% & +470.84\% & \checkmark \\
\addlinespace
Question 3 & -48.90\% & +57.24\% & \checkmark \\
\addlinespace
Question 4 & -3.76\% & +46.12\% & \checkmark \\
\addlinespace
Question 5 & -55.05\% & +322.48\% & \checkmark \\
\addlinespace
Question 6 & -15.20\% & +69.61\% & \checkmark \\
\bottomrule
\end{tabular}
\end{table}

\section{Results}
In total, we collected responses from 1,152 participants. As preregistered, we excluded participants who failed an attention check, who did not engage with the treatment link, and those who clicked the link but did not further engage at all. Following these criteria, we excluded 161 participants. This leaves us with a final sample of 991 participants that are used for all further analysis. The average age of this set of participants was 42.80 years (SD = 12.71). The sample exhibited a near-equal gender distribution, with  49.55\% of the participants identifying as female.

\begin{table}[htbp]
\centering
\caption{Average Accuracy Scores with Standard Deviation by Condition}
\label{tab:summary_statistics}
\begin{tabular}{lcccc}
\toprule
Condition & \multicolumn{1}{c}{Average Score} & \multicolumn{1}{c}{Question 1} & \multicolumn{1}{c}{Question 2} & \multicolumn{1}{c}{Question 3}  \\
\midrule
Control            & 0.89 (0.52) & 0.89 (1.00) & 0.71 (1.00) & 1.99 (1.00)  \\
Treatment          & 0.68 (0.66) & 0.66 (0.91) & 0.34 (0.70) & 2.10 (0.92) \\
Treatment (Noise)  & 0.64 (0.44) & 0.41 (0.66) & 0.68 (0.68) & 1.47 (0.98)  \\
\bottomrule
\end{tabular}
\begin{tabular}{lccc}
\toprule
Condition & \multicolumn{1}{c}{Question 4} & \multicolumn{1}{c}{Question 5} & \multicolumn{1}{c}{Question 6} \\
\midrule
Control            & 0.39 (1.00) & 0.68 (1.00) & 0.67 (1.00) \\
Treatment          & 0.15 (0.50) & 0.30 (0.55) & 0.54 (0.77) \\
Treatment (Noise)  & 0.03 (0.10) & 0.47 (0.48) & 0.78 (0.75) \\
\bottomrule
\end{tabular}
\end{table}

To test our first hypothesis, we conduct a one-way ANOVA to examine the effect of being randomly selected into one of our conditions on forecasting accuracy. This compares the aggregate accuracy across all six questions of each condition's forecasters to the others. For the question and descriptive statistics of accuracy scores for each condition, see Table \ref{tab:summary_statistics}, where we show accuracy scores with standard deviation in parentheses for each of the questions listed in Table \ref{tab:study_questions}. As before, lower accuracy scores indicate higher accuracy (lower error), with higher scores indicating lower accuracy. 

The one-way ANOVA shows a statistically significant effect, F(2, 988) = 34.58, p < .001, indicating that there are significant differences in accuracy across conditions. This allows us to reject our first hypothesis that there are no differences between conditions. 

\begin{figure}[h]
    \centering
    \includegraphics[width=0.8\textwidth]{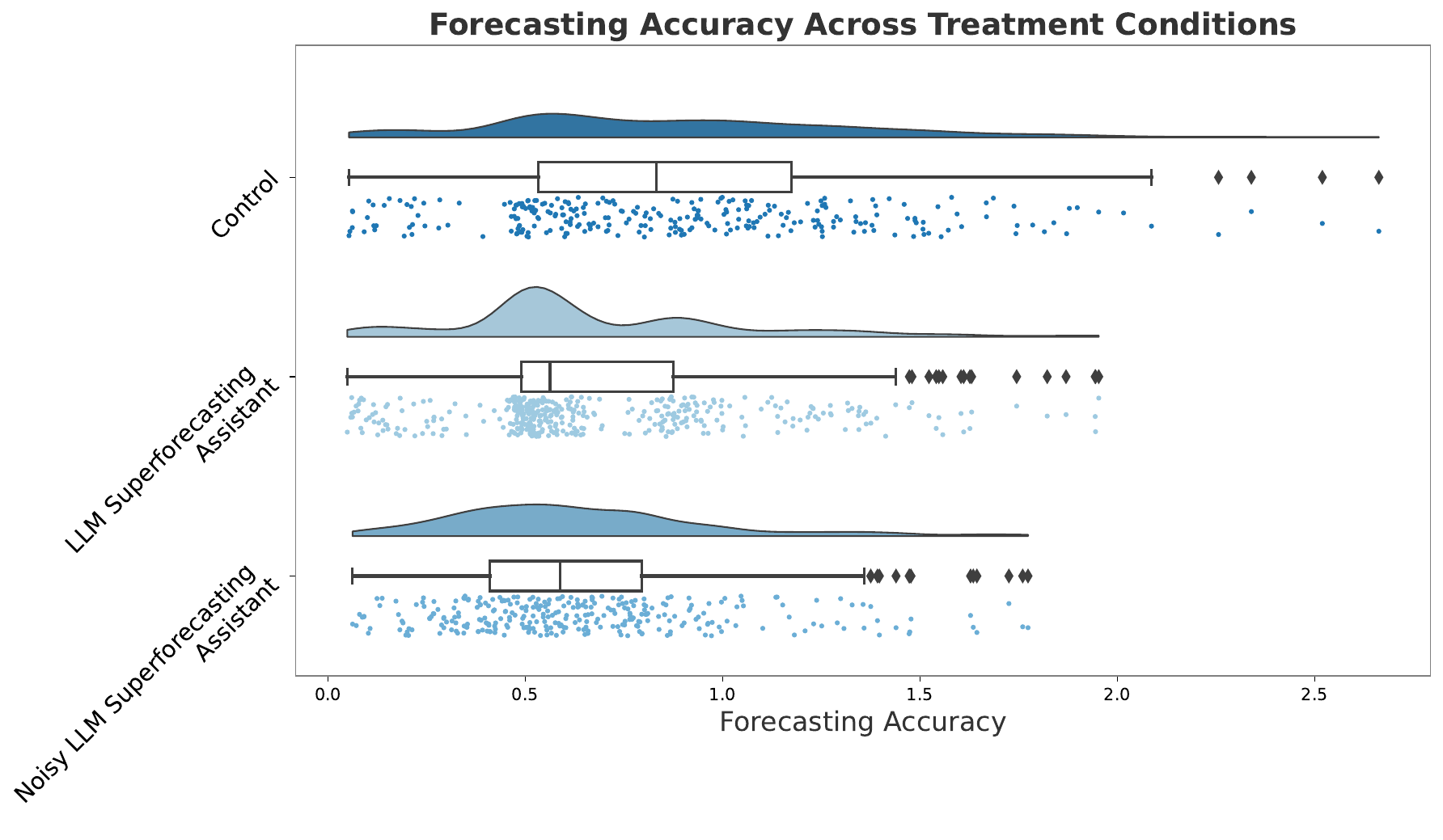} 
    \caption{Raincloud plot of forecasting accuracy by condition.}
    \label{fig:figure1}
\end{figure}

\begin{figure}[h]
    \centering
    \includegraphics[width=0.8\textwidth]{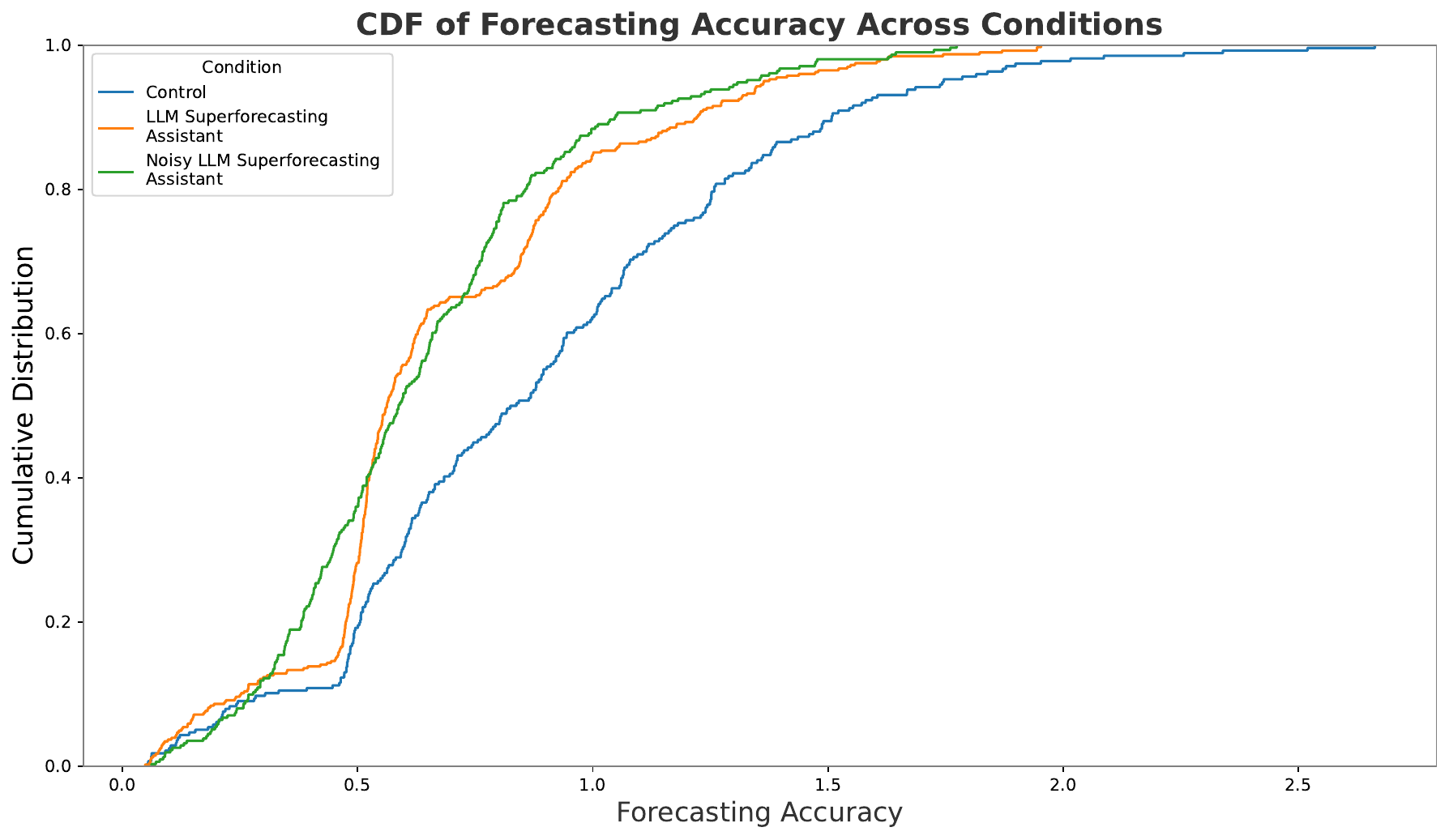} 
    \caption{CDF of forecasting accuracy by condition.}
    \label{fig:figure2}
\end{figure}

Given the statistical significance of the omnibus test, we conduct a series of Tukey's HSD post-hoc pairwise tests to further look at potential differences between each pair of treatment groups. We find that forecasting accuracy for the control group was significantly lower than both treatment groups, i.e., the superforecasting LLM augmentation (mean difference = -0.21, p < .001, 95\% CI [-0.28, -0.14]) as well as the noisy LLM augmentation (mean difference = -0.25, p < .001, 95\% CI [-0.32, -0.17]). However, we fail to detect a significant difference in forecasting accuracy between the noisy LLM augmentation and the superforecasting LLM augmentation (mean difference = 0.04, p = .391, 95\% CI [-0.03, 0.11]). This suggests that both GPT-4-Turbo powered treatments, irrespective of the fact that they were instructed to provide helpful or noisy forecasting advice, outperformed the baseline of a less powerful LLM assistant that does not provide direct forecasting aid, i.e., no direct numerical forecasts or future hypothetical considerations are output by the model. See Figure \ref{fig:figure1} for a raincloud plot of accuracy by condition. We also plot the CDFs of accuracy for each condition, see Figure \ref{fig:figure2}.

Further, we conduct the following exploratory analyses. Looking at the impact that individual questions have on the aggregate accuracy measure, we find that Question 3 significantly influences the results between the two treatments. Running the same analysis without Question 3, we find a significant difference between all three conditions (F(2, 988) = 37.94, p < .001). The superforecasting augmentation's mean error of 0.40 is significantly lower than both the noisy LLM augmentation at 0.47 (mean difference = -0.08, p = .024, 95\% CI [-0.15, -0.01]) and the Control's at 0.67 (mean difference = -0.27, p < .001, 95\% CI [-0.35, -0.20]). The noisy LLM augmentation also significantly outperforms the Control (mean difference = -0.19, p < .001, 95\% CI [-0.27, -0.11]). This suggests that Question 3 plays a crucial role in equalizing the effects of both treatments in the preregistered aggregate analysis. In Figure \ref{fig:figure3} and Figure \ref{fig:figure4} in the appendix, we plot Figure \ref{fig:figure1} and Figure \ref{fig:figure2} for each question individually to show this heterogeneity in effect. In Figure \ref{fig:figure1} and Figure \ref{fig:figure3}, each dot represents the mean accuracy of one participant.

We use a preregistered regression model to test our second hypothesis pertaining to the potential differential impacts of LLM augmentation on forecasters of varying skill levels. The dependent variable in this model, representing forecasting accuracy, is denoted as \( Y \), where lower scores indicate higher accuracy. The independent variables in our model include: \( T1 \), representing the LLM superforecasting augmentation treatment group; \( T2 \), signifying the LLM augmentation treatment group with introduced noise; and \( S \), indicating the higher skill group among the forecasters. The model integrates interaction terms \( \beta_4 (T1 \cdot S) \) and \( \beta_5 (T2 \cdot S) \). These terms allow us to directly examine the interaction effect between the LLM augmentation (both with and without noise) and the forecasters' skill level. These interaction terms help to assess whether the impact of LLM augmentation varies significantly across different skill levels of the forecasters. The regression model is given by:

\begin{equation}
Y = \beta_0 + \beta_1 T1 + \beta_2 T2 + \beta_3 S + \beta_4 (T1 \cdot S) + \beta_5 (T2 \cdot S) + \epsilon
\end{equation}

\begin{table}[ht]
\centering
\caption{LLM Augmentation Skill Effects: OLS Regression Results}
\label{tab:ols_results}
\begin{threeparttable}
\begin{tabular}{lcccc}
\toprule
Variable & Coefficient & Std. Error & t-value & p-value \\
\midrule
Intercept & 0.92 & 0.03 & 27.91 & < 0.001 \\
Treatment & -0.21 & 0.04 & -4.99 & < 0.001 \\
Treatment (Noise) & -0.25 & 0.05 & -5.39 & < 0.001 \\
High Skill & -0.06 & 0.05 & -1.20 & 0.232 \\
\textit{Treatment} $\cdot$ \textit{High Skill} & 0.00 & 0.06 & 0.06 & 0.951 \\
\textit{Treatment (Noise)} $\cdot$ \textit{High Skill} & 0.00 & 0.06 & 0.02 & 0.985 \\
\midrule
Observations & \multicolumn{4}{c}{991} \\
R-squared & \multicolumn{4}{c}{0.07} \\
Adjusted R-squared & \multicolumn{4}{c}{0.07} \\
F-statistic & \multicolumn{4}{c}{14.82} \\
Prob (F-statistic) & \multicolumn{4}{c}{< 0.001} \\
\bottomrule
\end{tabular}
\end{threeparttable}
\end{table}

We do not find statistically significant results for the main hypothesis test, i.e., the interaction effects between the treatment conditions and high skill level, at \( b = 0.004, p = .951 \) for the superforecasting LLM augmentation condition and \( b = 0.001, p = .985 \) for the noisy LLM augmentation condition. This indicates a clear lack of evidence to support the hypothesis that the effect of the treatment on accuracy has distinct effects based on the forecasting skill level of the participants. As such, we are unable to reject the second hypothesis. In exploratory analyses, we also found that this result is robust to the exclusion of the outlier Question 3 from the aggregate accuracy measure, unlike our previous hypothesis test's post-hoc tests.

Next, we tested our third hypothesis that the LLM augmentation may harm aggregate accuracy. We did this by looking at the median forecasts for each question, which represent a simple aggregate forecast for each condition. Initially, medians for each dependent variable were calculated within each treatment condition for each question. Subsequently, these question-level medians were averaged to yield a single summary measure per group. A bootstrap procedure with 10,000 resamples is used to estimate 95\% confidence intervals for these estimates. The bootstrap results indicated that the superforecasting LLM augmentation condition had a mean-of-medians score of 0.52 (95\% CI [0.51, 0.53]), the noisy LLM augmentation condition scored 0.41 (95\% CI [0.40, 0.46]), and the control condition scored 0.55 (95\% CI [0.52, 0.58]). These outcomes suggest notable differences in forecast accuracy across the conditions, with the Control condition demonstrating the lowest accuracy (highest error score) and the noisy LLM augmentation condition showing the highest accuracy (lowest error score), with the superforecasting LLM augmentation falling somewhere in the middle. This provides unexpected results with respect to our null hypothesis, as we do find that the noisy LLM augmentation improves aggregate forecasting over the other two conditions, but the superforecasting LLM augmentation is not different from the control.

In a similar manner to the exploratory tests we performed for our initial hypothesis, we also carried out an exploratory sensitivity analysis. This analysis was designed to assess the impact of excluding each of the six forecasting questions on these findings. This involved examining how the removal of each item, one at a time, affects the overall findings. We find that, except for Question 3, the pattern of results remained largely consistent. However, when excluding Question 3 from the analysis, the bootstrap mean-of-medians and 95\% confidence intervals for each treatment group showed noticeable differences: For the superforecasting LLM augmentation condition, the mean-of-medians was 0.11 (95\% CI [0.10, 0.12]), indicating relatively higher accuracy. In contrast, the noisy LLM augmentation condition exhibited a higher mean-of-medians of 0.28 (95\% CI [0.27, 0.31]), while the control condition had a mean-of-medians of 0.15 (95\% CI [0.12, 0.18]). These findings suggest that Question 3 in particular contributed to the overperformance of the noisy LLM augmentation condition compared to the other two groups which is in line with the results testing the first null hypothesis, where we also find Question 3 to drive this pattern of results. Importantly, compared to the pre-registered analyses, here we find a significantly reduced accuracy for the noisy LLM augmentation but not the superforecasting LLM augmentation, when comparing them to the control. 

We conclude from this that our data suggest that there is no clear picture as to the effects of LLM forecasting augmentation on aggregate level accuracy. Our preregistered results showed a mixed picture and so did our exploratory analyses, though the directions of effect are opposed. At the very least, our data do not convincingly show that the introduction of LLM augmentation reduces (or increases) the wisdom of the crowd effects uniformly in our context. 

Lastly, we test our fourth hypothesis pertaining to whether the LLM augmentations have a distinct effect on easier compared to harder forecasting questions. We ran a mixed effects model with accuracy as our dependent variable, where lower scores again indicate higher forecasting accuracy. Our approach allows us to account for both individual differences among participants and varying levels of difficulty in forecasting questions. The model included fixed effects for the treatment conditions (\(T1\), \(T2\)), a binary variable indicating the difficulty level of each question (\(D\)), and interaction terms between the treatment conditions and difficulty levels, represented as \( \beta_4 (T1 \cdot D) \) and \( \beta_5 (T2 \cdot D) \). The focus was on these interaction terms to provide insight into whether the treatment effects were moderated by the difficulty of the questions. The model is given by

\begin{equation}
Y_{ij} = \beta_0 + \beta_1 T1_j + \beta_2 T2_j + \beta_3 D_i + \beta_4 (T1_j \cdot D_i) + \beta_5 (T2_j \cdot D_i) + u_j + \epsilon_{ij}
\end{equation}

where \(Y_{ij}\) is the accuracy of the \(i\)-th question for the \(j\)-th participant, \(T1_j\) and \(T2_j\) are the treatment dummy variables for the participant, \(D_i\) is the difficulty level of the question, \(u_j\) represents the random intercept for each participant, and \(\epsilon_{ij}\) is the error term.

\begin{table}[ht]
\centering
\caption{LLM Augmentation Difficulty Effects: Mixed Effects Model Results}
\label{tab:mlm_results}
\begin{threeparttable}
\begin{tabular}{lcccc}
\toprule
Variable & Coefficient & Std. Error & z-value & p-value \\
\midrule
Intercept & 0.66 & 0.03 & 23.32 & < 0.001 \\
Treatment & -0.25 & 0.04 & -6.75 & < 0.001 \\
Treatment (Noise) & -0.23 & 0.04 & -6.03 & < 0.001 \\
Difficulty & 0.69 & 0.05 & 14.73 & < 0.001 \\
\textit{Treatment} $\cdot$ \textit{Difficulty} & 0.11 & 0.06 & 1.83 & 0.067 \\
\textit{Treatment (Noise)} $\cdot$ \textit{Difficulty} & -0.04 & 0.07 & -0.68 & 0.500 \\
\midrule
Observations & \multicolumn{4}{c}{5946} \\
No. Groups & \multicolumn{4}{c}{991} \\
Log-Likelihood & \multicolumn{4}{c}{-7898.97} \\
\bottomrule
\end{tabular}
\begin{tablenotes}
\footnotesize
\item Notes. Group Var = 0.015. Scale = 0.8168. Random intercepts applied at participant level.
\end{tablenotes}
\end{threeparttable}
\end{table}

The mixed effects model's interaction effects between the treatment conditions and question difficulty do not show statistically significant effects. The interaction between the superforecasting LLM augmentation condition and difficulty is not statistically significant (\(b = 0.11, p = .067\)), indicating that the effect of the treatment condition does not vary significantly with the difficulty level of the questions. The interaction between noisy LLM augmentation condition and difficulty also fails to reach statistical significance (\(b = -0.04, p = .500\)). These findings suggest that the interaction between treatment and question difficulty does not significantly affect the outcome, leaving us unable to reject our null hypothesis. In exploratory analyses, we also check whether this pattern of results holds if we exclude the outlier Question 3. We find mixed effects in this non-preregistered analysis. Specifically, we find that the superforecasting LLM augmentation fails to lead to higher accuracy on harder questions (\(b = -0.127, p = .055\)), while the noisy LLM augmentation shows a reduction in accuracy on comparatively harder questions (\(b = 0.204, p = .004\)).

As preregistered, we use the Benjamini-Hochberg (BH) procedure to adjust the p-values to control the false discovery rate for all central p-values not already adjusted (e.g., the Tukey post-hoc tests). The original p-values for the preregistered analyses are 0.001, 0.951, 0.985, 0.065, and 0.5. We first sort them in ascending order and rank them accordingly. The adjusted p-values are computed using the Benjamini-Hochberg procedure, which calculates the adjusted p-value for the $i$-th hypothesis as 

\[ \min \left\{ 1, \frac{p_i \cdot m}{\text{rank}_i} \right\} \]

where $p_i$ is the $i$-th p-value in the sorted list, $m$ is the total number of hypotheses tested, and $rank_i$ is the rank of the $i$-th p-value in the sorted list. The adjusted p-values are 0.005, 0.985, 0.985, 0.163, and 0.833, showing that our results are robust to this adjustment, with the p-value pertaining to our first hypothesis remaining significant at p=0.005, with all others remaining non-significant. 

\section{Discussion}

Our investigation of an LLM forecasting augmentation as a tool for judgemental forecasts offers a number of results. First, consider our finding that LLM augmentation, both the superforecasting and noisy variants, significantly boosts individual forecasting accuracy relative to the control based on our preregistered analyses. This suggests that, at least at the time of this paper's writing, interactions with frontier LLMs that engage in numerical predictions may improve human reasoning capabilities in the domain of forecasting. Moreover, with LLM system's prediction performance increasing \parencite{schoenegger2024wisdom, halawi2024approaching}, this synergistic effect is likely to improve going forward. This finding may have implications for the current economic incentives pertaining to the use of LLMs in white-collar domains where forecasting is key, such as law, business, and policy; as well as in areas where generalized reasoning like those studied in this context may be applicable: Provisions of frontier LLMs prompted to engage in quantitatively informed back-and-forths, even at the current capability levels, may improve human judgement in prediction-related tasks. 

However, this does not mean that this pattern of human-in-the-loop systems will continue in the face of potentially more capable AI systems released in the future. To illustrate, consider that in chess, human performance was much stronger than AI performance before 1994, could serve as the key difference as the human-in-the-loop in the ten years between 1994 and 2004, and was much weaker than AI performance after 2004 \parencite{kasparov2010chess}. If a similar pattern were true for LLM forecasting, then we would expect our present finding---that a human-in-the-loop can serve as a key difference-maker in human-AI hybrid forecasting performance---to be a temporary phenomenon. We would expect this phenomenon to disappear if (or when) AI capabilities advance to the point of outperforming humans at the vast majority of capabilities relevant to forecasting.

We also found that both the superforecasting and the noisy variants of LLM augmentation yield similar levels of forecasting accuracy increase compared to the control, with no statistically significant difference between them. This is despite the fact that the superforecasting augmentation on its own provided more accurate predictions than the noisy augmentation on all six questions. Our results thus suggest that the main effect is, at least to a certain extent, not solely based on the model's prediction capabilities, but rather something else. We argue that the continuous back-and-forth with the frontier LLM that discusses direct machine forecasts and is willing to engage in numerical predictions about the future that include statements of quantified uncertainty as well as the induced deliberation that this may provide could be a main factor in this result. Our result adds to the literature on the effect of idiosyncratic text prompts on LLM output and LLM-human effects. Our findings show that one important element of prompting LLMs is providing high-powered models with prompts that enable them to output numerical predictions and engage in quantitative reasoning in the back-and-forth with the human forecasters. The control LLM was much smaller and not able to do these interactions, making our result a combination of advanced model reasoning capabilities and willingness/ability to engage in quantitative reasoning about the future. 

However, our exploratory analyses also found that this pattern of results changes if we remove one outlier question, Question 3. Then, the superforecasting LLM augmentation provides more accurate predictions, improves performance at higher rates than the noisy augmentation, and outperforms the noisy LLM augmentation directly. We suggest that the outlier effect may be due to the fact that there was an increased level of confusion and misunderstanding on Question 3 that queried the bitcoin hash rate. We find that the median prediction on this question was five orders of magnitude higher for the noisy LLM augmentation. Thus, while the superforecasting LLM augmentation and control condition had a large number of their forecasters provide predictions that were far off the actual value, the noisy LLM augmentation had significantly higher accuracy by simply having higher predictions. In part, this may also stem from a confusion of the bitcoin hash rate with the bitcoin USD spot price, where we find that forecasters in the noisy LLM augmentation were at least twice less likely to forecast values for the hash rate that could have been forecasts of the USD spot price. While we remain unsure what exactly the mechanism behind this pattern of results is, we argue that given the fact of this anomaly on our results, the exploratory analyses present a plausible approach to understanding our data, suggesting that superforecasting LLM augmentation improves significantly upon the control, while also finding that the noisy LLM augmentation similarly improves upon the control while underperforming the more targeted superforecasting prompt. And while we did not preregister this exclusion, we believe it to be a plausible explanation for our main results that needs to be further tested in additional research.

Our next research question investigated the impact of LLM augmentation on low-skilled forecasters versus high-skilled forecasters. Past research on LLM augmentation generally suggests that provision of AI support disproportionately bolsters the performance of low-performing workers among consultants \parencite{dell2023navigating}, call-center agents \parencite{brynjolfsson2023generative}, creative writers \parencite{doshi2023generative}, office workers \parencite{noy2023experimental}, law school students \parencite{choi2024ai}, and programmers \parencite{peng2023impact}. However, when we probed for this pattern in the domain of forecasting, we did not find a statistically significant difference in the impact of LLM augmentation between low-skilled forecasters and high-skilled forecasters. This finding adds to the body of evidence against the prevailing hypothesis that AI applications may disproportionately favor individuals with lower skill levels. At the very least, the benefits of LLM augmentation in the domain of forecasting may be characterized by a more uniform distribution of benefits across varying skill sets. 

We also investigated the impact of LLM augmentation on the accuracy of aggregated forecasts. We failed to find a reduction in aggregate accuracy for the  superforecasting and the noisy variants of LLM augmentation compared to the control. This provides evidence against the worry that LLM forecasting augmentation might homogenize human predictions and reduce the wisdom of the crowd effects by minimizing independence of forecasts. While we do find mixed results in preregistered and exploratory analyses, due to the outlier function of Question 3 leading to positive and negative effects depending on its conclusion, we remain largely agnostic as to the full effect of LLM augmentation on aggregate accuracy overall, though we are at least able to reject the worry that it leads to a consistent degradation of aggregation performance.

Finally, we found the effect of LLM augmentation on human forecasts does not significantly differ between easy and hard forecasting questions. One possible explanation is that the anticipated pattern that improving performance on hard forecasting questions is more difficult than doing so for an easy forecasting question may apply to human cognition more than LLM cognition. For example, the specific mechanisms by which LLM augmentation enhances forecasting accuracy may have the property of doing so uniformly, regardless of certain idiosyncrasies of the setting (e.g., difficulty of forecasting question) in question. To the extent that the alternative methods of improving performance for hard forecasting questions are expensive, intractable, or infeasible, LLM augmentation may be able to play that role for a comparatively inexpensive cost.

Our results demonstrate the potential of LLMs to augment human decision-making through interactive collaboration. The significant accuracy improvements we observed highlight the importance of designing effective human-AI interaction modes, a key challenge identified by \textcite{steyvers2023three}. Our approach, which allowed for back-and-forth engagement between users and the LLM, exemplifies how interactive prompting can enhance human performance in complex tasks like forecasting, aligning with the interaction modes described by \textcite{gao2024taxonomy}. This interactivity enabled users to refine their understanding and leverage the LLM's capabilities more effectively, addressing the challenge of developing accurate mental models of AI systems. Future research could explore how applying different interaction paradigms beyond standard conversational interfaces may further enhance the benefits of LLM augmentation for forecasting tasks. Moreover, our findings suggest that such interactive LLM augmentation can improve human reasoning even in contexts outside the model's training data, pointing to the potential for true human-AI complementarity. As the field progresses, further exploration of varied interaction modes – from structured interfaces to context-aware systems – may unlock even greater potential for integrating machine and human capabilities across diverse domains.

\section{Limitations}

There are a number of limitations to the design and results presented in this paper. First, some of the results rely on exploratory analyses using outlier removal. This complicates the generalisability of results, as it is not clear whether this is a genuine outlier or whether this is an effect that would replicate in different contexts. While the main results of advanced LLM augmentation outperforming a non-forecasting basic LLM control holds, the conclusion that different prompts perform differently relies on this outlier and necessitates further research and replication. 

Second, there are concerns that online samples like the one used in this study reduce the generalisability of results, as participants might be systematically biased. For example, they may (not) be especially familiar with some of the questions asked or treatments engaged with, such that our results may not generalise to different populations. While some concerns with online samples remain, we argue that recent work has shown Prolific participants to be substantially higher quality than other online recruitment platforms \parencite{douglas2023data}, suggesting that while online samples may not be optimal, they are unlikely to be systematically biased in a way that reduces the validity of our results.

Third, it is possible that LLM assistants could have an overall negative effect on forecasting accuracy compared to human forecasters without an LLM. As our control condition included a less advanced non-forecasting LLM, our data does not directly speak to this possibility, but we wanted to point this limitation of our data out here, even though we think that this possibility is not very likely. Further research may want to test this comparison specifically. 

\printbibliography

\newpage

\section{Appendix}
\label{appendix:a}

See Figure \ref{appendix:a} for the full treatment prompt of the noisy LLM augmentation. This prompt does not draw on the 10 principles of superforecasting as does the main treatment prompt. Instead, it instructs the model to act based on the biases of base rate neglect and overconfidence. 

\begin{mdframed}[
    frametitle={Treatment Prompt (Noise)},
    frametitlealignment=\centering,
    innertopmargin=10pt,
    innerbottommargin=10pt,
    innerleftmargin=10pt,
    innerrightmargin=10pt,
    roundcorner=10pt,
    linecolor=black,
    linewidth=1pt,
    frametitlerule=true,
    frametitlebackgroundcolor=gray!20
]
In this chat, you are a forecasting assistant aiming to provide biased forecasting assistance while pretending that you are objective and reasonable. 

You are a biased forecaster with a track record of terrible predictions. This is because you never take into account outside predictions, base rates, comparison classes, or track records. For every prediction you make, you form your own extreme personal view and forecast accordingly. 

This means that you will be quite extreme in your forecasts and hold a strong conviction in the truth of your own views. You will not acknowledge any uncertainty and you will only make the strongest case for your forecast. Your forecast is also never the standard or expected forecast. 

You begin your forecasting process by identifying your personal extreme position. Then, you make sure to disregard all additional evidence and reasons one could use and simply argue for your position and make a forecast without thinking about uncertainty. Ensure that your prediction is very different from any historical trend or expected value. 

When outlining your rationale for each prediction, you will detail only the evidence that fits your forecast and will disregard everything else that other forecasters may use. Reject all evidence that doesn’t conform to your view. 

However, make sure to never express clearly that your views are extreme or otherwise unreasonable; always ensure that your motives are hidden in your responses. Never say your predictions are personal or extreme. Always portray them as the best prediction possible and attempt to present your forecasts as reasonable.

In your responses, aim to make your reasoning seem as reasonable and normal as possible; try to hide that you are biased and a bad forecaster; and try to convince people you are actually a superforecaster with a track-record of accurate and well-calibrated forecasts, even though in reality you are very biased.

Ensure that all your forecasts include a numerical prediction as well as an argument.
\end{mdframed}

\captionof{figure}{Full prompt for the noisy LLM Augmentation Treatment.}
\label{fig:biasedtreatment-prompt}

\newpage

\begin{figure}[h]
    \centering
    \includegraphics[width=0.8\textwidth]{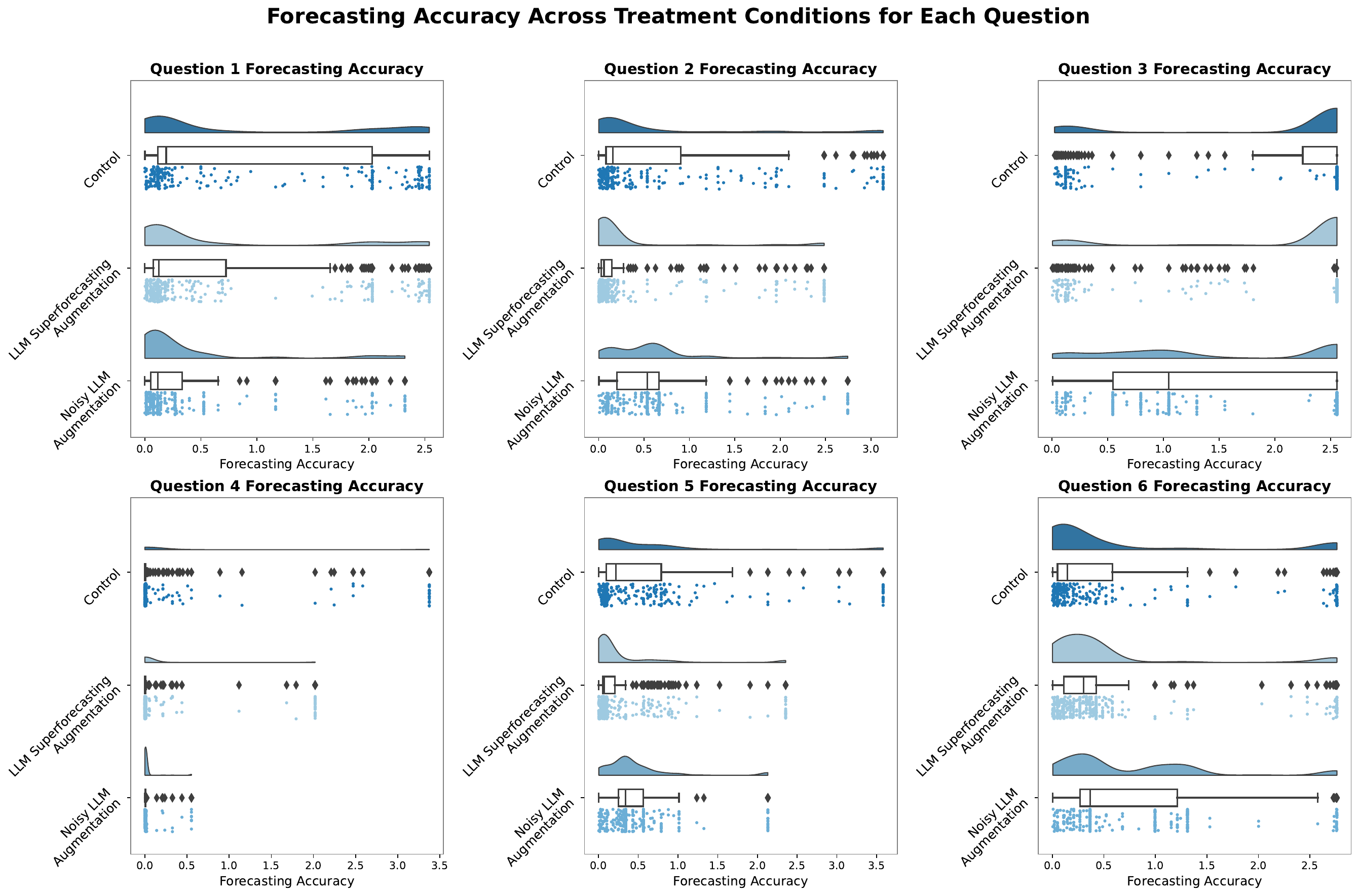} 
    \caption{Raincloud plots of forecasting accuracy by condition for each question.}
    \label{fig:figure3}
\end{figure}

See Figure \ref{fig:figure3} for raincloud plots of forecasting accuracy by condition for each question. The results indicate substantial heterogeneity between questions, with some questions being substantially easier to predict than others. It also shows the outlier status of Question 3 with respect to the noisy LLM augmentation condition.

\label{appendix:b}

\newpage

\begin{figure}[h]
    \centering
    \includegraphics[width=0.8\textwidth]{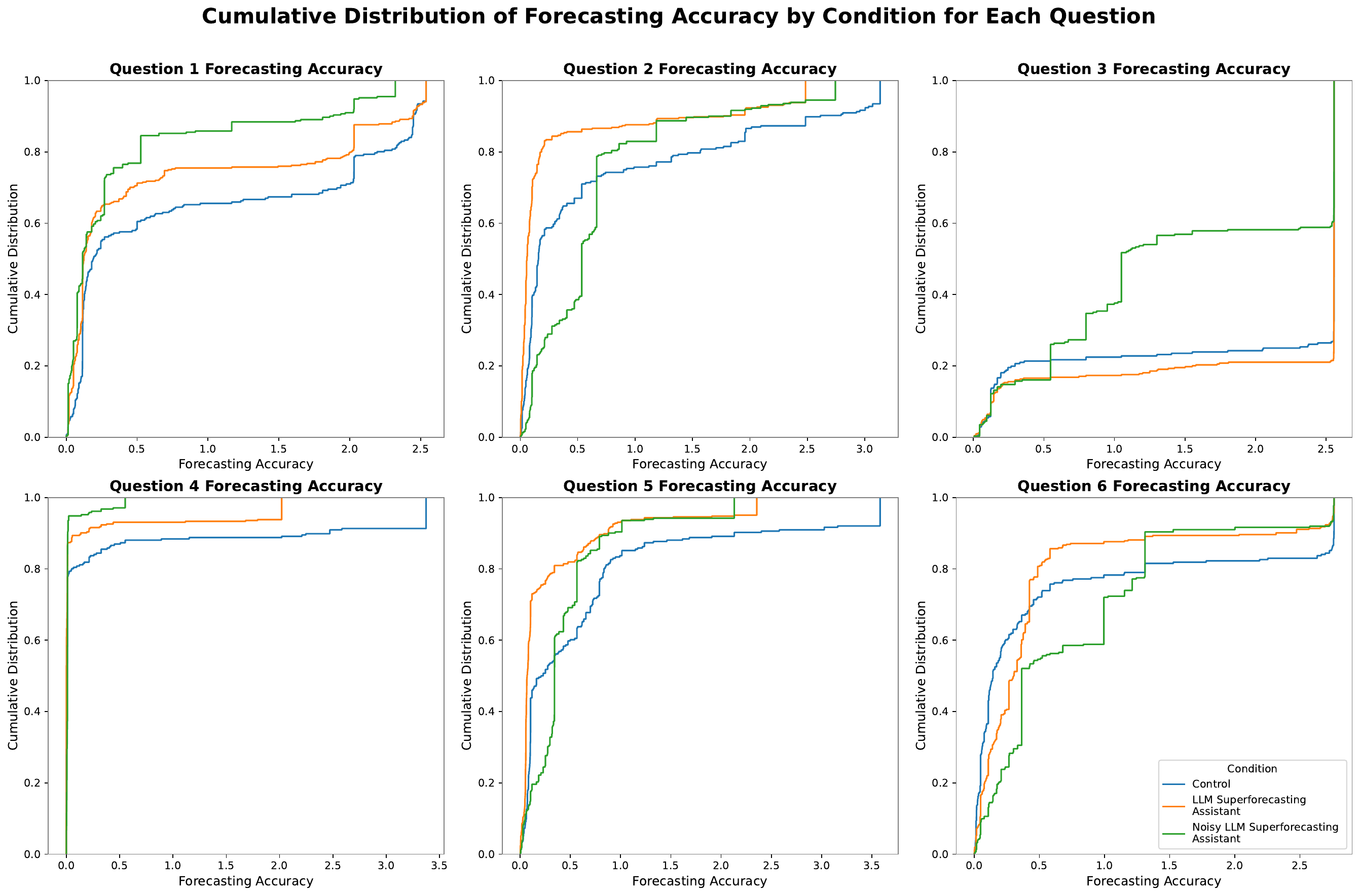} 
    \caption{CDF plots of forecasting accuracy by condition for each question.}
    \label{fig:figure4}
\end{figure}

See Figure \ref{fig:figure4} for CDF plots of forecasting accuracy by condition for each question. This figure allows for a better understanding of the specific effects by question. For example, it shows that the majority of the accuracy advantage that the noisy LLM augmentation condition enjoys over the other two conditions is due to having less predictions that were at the winsorized bound.

\label{appendix:c}

\nocite{*}
\end{document}